\documentclass[useAMS,usenatbib]{mn2e}
\usepackage{graphicx}
\usepackage{color}

\newcommand{\be}{\begin{equation}}
\newcommand{\ee}{\end{equation}}
\newcommand{\ba}{\begin{eqnarray}}
\newcommand{\ea}{\end{eqnarray}}

\title[Shortterm variations of AM CVns]{The influence of short term variations in AM CVn systems on LISA
measurements}
\author[Stroeer et al.]{A. Stroeer$^{1}$\thanks{E-mail:
Alexander.Stroeer@nasa.gov, now at CRESST, Department of Astronomy, University of Maryland, College Park, Maryland 20742 and Laboratory for Gravitational Physics, Goddard Space Flight Center, Greenbelt, Maryland 20771} and G. Nelemans$^{2}$\footnotemark[1]\thanks{We acknowledge contributions to research and paper editing by Alberto Vecchio.}\\
$^{1}$School of Physics and Astronomy,
University of Birmingham,
Edgbaston,
Birmingham B15 2TT,
United Kingdom\\
$^{2}$Department of Astrophysics, IMAPP,
Radboud University Nijmegen,
Heyendaalseweg 135,
NL-6525 AJ Nijmegen}
\begin{document}

\date{}

\pagerange{\pageref{firstpage}--\pageref{lastpage}} \pubyear{2009}

\maketitle

\label{firstpage}

\begin{abstract}
We study the effect of short term variations of the evolution of AM
CVn systems on their gravitational wave emissions and in particular
LISA observations. We model the systems according to their equilibrium
mass-transfer evolution as driven by gravitational wave emission and
tidal interaction, and determine their reaction to a sudden perturbation
of the system. This is inspired by the suggestion to explain the
orbital period evolution of the ultra-compact binary systems V407 Vul
and RX-J0806+1527 by non-equilibrium mass transfer.  The
characteristics of the emitted gravitational wave signal are deduced
from a Taylor expansion of a Newtonian quadrupolar emission model, and
the changes in signal structure as visible to the LISA mission are determined. We show that
short term variations can significantly change the higher order terms
in the expansion, and thus lead to spurious (non) detection of
frequency derivatives. This may hamper the estimation of the parameters of the system, in particular their masses and
distances. However, we find that overall detection is still secured as signals still can be described by general templates. 
We conclude that a
better modelling of the effects of short term variations is needed to
prepare the community for
astrophysical evaluations of real gravitational wave data of AM CVn systems.
\end{abstract}


\section{Introduction}

The Laser Interferometer Space Antenna (LISA) is an ESA/NASA
space-based gravitational wave (GW) laser interferometer designed to
observe a wide range of sources in the frequency range $\sim 10^{-5}$
- 1 Hz \citep{Benderr1998a}. At the moment LISA is in its design stage
with an open launch window of 2018+. One of the important classes of
objects are Galactic compact object binary systems, in particular
double white dwarf binaries
\citep{1987ApJ...323..129E,1988ApSS.145....1L,1990ApJ...360...75H}. LISA
will identify several thousand new binaries -- with the key feature
of not being biased against short orbital periods -- and will
therefore contribute in a significant way to unveil the complex
physical interactions in compact object binaries
\citep[e.g.][]{2004MNRAS.349..181N,Stroeer2005a,2007CQGra..24..513B}. A
particular class of objects are the mass-transferring AM CVn binaries
\citep{Nel2005a}. In these systems the orbital evolution of the system
is not only determined by gravitational wave emission (and possible
tidal interaction), but also by the mass redistribution in the
system. {A number of the currently known AM CVn systems are expected to serve as 
verification sources for LISA \citep{Stroeer2006a, 2007ApJ...666.1174R}.}

In an earlier paper \citep{Stroeer2005a} we showed that in
order to determine the system parameters, either the exact evolution of
the gravitational wave frequency up to second order in a series
expansions needs to be determined, or complementary electro-magnetic
observations are needed. Here we revisit this result, focusing on
possible short term variations in the equilibrium mass-transfer rate,
as observed in many interacting binaries, and as have been proposed as
explanation for the observed period decrease in the ultra-compact
binary systems V407 Vul and RX~J0806+1527 \citep[][and references
therein]{marsh2005a,2005MNRAS.357.1306B}. In addition, especially
early in the evolution of AM CVn systems, helium novae are expected
\citep{2007ApJ...662L..95B}, leading to a sudden perturbation of the
system. 

In Sect.~\ref{sec:mod} we describe the basic model, the way we
perturb the system and the calculation of the resulting gravitational
wave signal. In Sec.~\ref{sec:dem} we discuss the results for one
fiducial AM CVn system after which we conclude in Sec.~\ref{sec:disc}.

\section{Model}
\label{sec:mod}

\subsection{Binary evolution and perturbation}\label{binev}

In this section we briefly review the key elements of the evolution of
an AM CVn system, and the model adopted to characterize short period
perturbations of the mass transfer rate. Our formalism is based on
\citet{Marsh2004a} to which we refer the reader for more details.

The progenitor stage of an AM CVn system is characterized by the loss
of orbital angular momentum due to the emission of gravitational
waves. The lost orbital angular momentum drives the system to shorter
orbital periods and smaller orbital separations. If the orbital
separation is sufficiently small for the radius of the secondary star
to overfill its Roche lobe, mass is transferred onto the primary star
(accretor) through Roche lobe overflow. The accretor responds by
spinning up, removing angular momentum from
the orbit at an even higher rate. However, tidal interaction may force
the accretor into co-rotation with the orbit, effectively returning
the lost angular momentum. For stable systems, redistribution of mass
in the system allows the orbit to expand, even at decreasing angular
momentum. The binary thus reaches a ``turning point''.

The binary under consideration is characterized by the masses and
radii of the primary (more massive and accreting) and secondary (less
massive and mass transferring) star, $M_{1,2}$ and $R_{1,2}$
respectively. The binary orbital separation $a$ and its period $P$
evolve due to the change of orbital angular momentum $J_\mathrm{o}$
according to \citep[e.g.][]{Marsh2004a}
\be
\dot{J_\mathrm{o}} = \dot{J}_\mathrm{GR} + \sqrt{G\,M_1\,R_\mathrm{h}}\,\dot{M}_2 + \frac{k M_1 R_1^2}{\tau_\mathrm{S}}\,(\Omega_\mathrm{s} - \Omega_\mathrm{o})\,;
\ee
the three terms on the right-hand side of the equation
represent the loss of angular momentum due to gravitational {wave}
radiation $\dot{J}_\mathrm{GR}$, mass loss at rate $\dot{M}_2$, and
{tidal} coupling, respectively. The latter
contribution is parametrized as a function of the synchronization
time-scale $\tau_\mathrm{S}$, {and} $R_\mathrm{h}$
identifies the radius around the accretor with the same angular
momentum as the transferred
matter~\citep{Ver1998a}. $\Omega_\mathrm{s}$ and $\Omega_\mathrm{o}$
are the angular frequencies of the accretor's spin and the orbit,
respectively and $k\approx 0.2$ \citep{Marsh2004a}.

We model the mass transfer rate $\dot{M}_2$ as an adiabatic response of the donor
\be\label{m:massloss}
\dot{M}_2=-f(M_1,M_2,a,R_2) \Delta^3\,,
\ee
where the function $f(M_1,M_2,a,R_2)$ is given by Eq. (10) of \cite{Marsh2004a} ({as} based on results from \citet{Web1977a}). 
The quantity $\Delta$ is the Roche lobe overflow factor, defined as
\be\label{eq:rlofd}
\Delta  = R_2 - R_\mathrm{L}\,,
\label{e:Delta}
\ee where $R_\mathrm{L}$ is the Roche lobe's radius {and a} function
of $a$ and $M_2/M_1$ only. We use a simple zero-temperature mass-radius relation as in \citet{Marsh2004a}, 
neglect the effects of a realistic donor structure, that generally 
gives rise to larger donor radii, i.e. lower mass-transfer rates and more 
complex mass transfer behaviour \citep[e.g.][]{2006ApJ...649L..99D, 2007MNRAS.381..525D}.

Mass transfer spins up the accretor. We therefore see a coupling of
mass transfer rates and the response of the accretor due to tidal
locking. This coupling can be expressed in a coupled differential
equation system of the evolution of $\Delta$ and the evolution of
$\Omega_s$
\ba
\frac{1}{2R_2}\frac{d\Delta}{dt}&=&-\frac{\dot{J}_\mathrm{GR}}{J_\mathrm{o}}-\frac{kM_1R_1^2}{\tau_\mathrm{S}J_\mathrm{o}}\,(\Omega_\mathrm{s} - \Omega_\mathrm{o}) + \frac{\dot{M}_2}{M_2} \times
\nonumber \\
&& \left(1+\frac{\zeta_2-\zeta_{r_L}}{2}-q-\sqrt{(1+q)\frac{R_h}{a}}\right)\,,\label{deq1}  \\
\frac{d\Omega_s}{dt}&=&\left(\lambda \Omega_s-\frac{\sqrt{GM_1R_h}}{k R_1^2}\right)\frac{\dot{M}_{2}}{M_1}-\frac{(\Omega_\mathrm{s} - \Omega_\mathrm{o})}{\tau_\mathrm{S}}\, \label{deq2}.
\ea
In the previous equation $q = M_2/M_1$ is the mass ratio and the
parameters $\zeta_{2}$, $\zeta_{r_L}$ and $\lambda$ are given,
respectively, by
\ba
\zeta_2 & = & \frac{d \log{R_2}}{d \log{M_2}}\,,
\\
\zeta_{r_L} & = & \frac{d \log{R_L/a}}{d \log{M_2}}\,,
\\
\lambda & = & 1+2\frac{d \log{R_1}}{d \log{M_1}}+\frac{d \log{k}}{d \log{M_1}}\,.
\ea

We started the modelling of our fiducial AM CVn system in the
progenitor phase and followed its evolution through the "turning
point" towards the stable AM CVn stage assuming strong tidal locking
($\tau_\mathrm{S}=0.2 yrs$ as favoured by \citet{Marsh2004a}, \textrm{red}{but note the discussion of the very large uncertainty in this value in their section 6}). Masses for
the primary and secondary star of our fiducial binary were defined in
the progenitor phase with $M_1=0.4M_{\odot}$ and
$M_2=0.2M_{\odot}$. 

We induce a perturbation by suddenly offsetting $\Delta$ {(through
a change in separation $a$ that in turn changes the Roche lobe
$R_L$)}, which directly impacts the mass transfer rates and spin evolution
of the accreting star (see Eqs.~\ref{m:massloss},\ref{deq1} and~\ref{deq2}). {As
  the mass transfer in the AM CVn systems is calculated implicitly
  from the system parameters this is the simplest self-consistent way
  of perturbing the system.} The
{expected} timescale $\tau_\mathrm{c}$ {for returning to
equilibrium after} such a perturbation is 
\citep{marsh2005a} \be
\tau_\mathrm{c}=\frac{1}{3}\frac{\Delta}{\dot{\Delta}} 
\ee 

We triggered the perturbation when the system reaches a gravitational
wave frequency of 8 mHz in its stable AM CVn branch.  We perform two
kinds of perturbations. 
{First we decrease the Roche lobe overflow factor, yielding
reduced mass transfer rates, possible e.g. when nova explosions eject
mass from the accreting star, widening the orbit. Second, we increase
the Roche lobe overflow yielding increased mass transfer rates. Such
an effect may be caused by suddenly expanding stellar atmospheres of
the mass donor, or decrease in orbit if the expanding nova shell
interacts with the secondary, removing orbital angular momentum.}  We
offset $\Delta$ {such as} to yield five times larger and five times smaller mass
transfer rates (see Eq.~\ref{m:massloss}). Effectively we disturb the
equilibrium rates of the AM CVn system which balance Eq.~\ref{deq1}
and Eq.~\ref{deq2}, and force the system to {evolve}
back to its equilibrium by itself. 

{A factor five as perturbation for mass transfer rates is a quite modest change. The separation only needs to be changed by much less than one per cent to achieve such an offset. Further, we restrict this case study to a system at GW frequency of 8 mHz, 
since we refer to 
novae as source for the perturbation. The occurrence of a nova depends on the ignition of the accreted 
layer, which happens much easier and more often at high mass transfer rates 
(i.e. short periods) than later \citep{2007ApJ...662L..95B}.}

\subsection{Gravitational wave signal}

We model the GW that is emitted from such a system at the leading Newtonian
quadrupole order, only keeping {the} $+$ polarisation and working in the
real domain. This yields for the GW strain in the solar reference
frame
\be
h_{GW}(t)=A(t) \cos(\Phi_{GW}(t))=A(t) \cos( 2\pi  t f_{GW}(t)),
\ee 
with 
\be
A(t)=-\frac{1}{D} \frac{G^2}{c^4} \frac{2M_1(t)M_2(t)}{a(t)}
\ee
the amplitude of the strain ($M_{1,2}(t)$ masses of  {the} stars in a
binary as affected by mass transfer, $D,a(t)$ distance to and varying
separation of the system respectively, $G,c$ Keplers constant and
{the} velocity of light respectively, $\Phi_{GW}(t)$ the phase and $f_{GW}$
the frequency of the gravitational wave). We {calculate} the period of the AM
CVn according to
\be
(P(t)/2\pi)^2=a(t)^3/G(M_1(t)+M_2(t))
\ee (Kepler's third law) to derive the gravitational wave frequency 
\be
f_{GW}(t)=2/P(t)
\ee 
(leading order quadrupolar radiation as found in circular orbits, we
drop the subscript $GW$ hereafter).

We do neglect the LISA instrumental response function in this
research, and thus remain in the solar reference frame through out. We
do note that this does not have repercussions for the conclusions, 
as the response function {does not change} the GW frequency {evolution} (it merely modulates it). In the remainder of this
paper we drop the explicit reference to the time dependence in the
parameters.

We expanded the evolving phase of gravitational wave emission from
{the} AM
CVn system as {a} Taylor series up to second order to determine
the 
 {influence of the perturbation to the GW signal.}
\be
\Phi_{GW} (t) = 2 \pi f_0 t + \pi f_0^{(1)} t^2 + \frac{\pi}{3}f_0^{(2)} t^3 
\ee
We defined with $f_0,f_0^{(1)},f_0^{(2)}$ spin-down parameters for the
system, here stating the gravitational wave frequency, its first and
second derivative in time at the origin of the series, {time nominally set to zero at
an instant after the sudden offset of $\Delta$}. $\Phi_{GW} (t)$ is
further the number of wave-cycles one expects to record at LISA within
a given observation time. For this the running time $t$ is replaced
with the total observation time $T$. In this view, the three {terms}
to $\Phi_{GW} (T)$ are contributing separately to the total number of
wave-cycles.

\section{Results}\label{sec:dem}
\begin{figure}
\resizebox{\hsize}{!}{\includegraphics{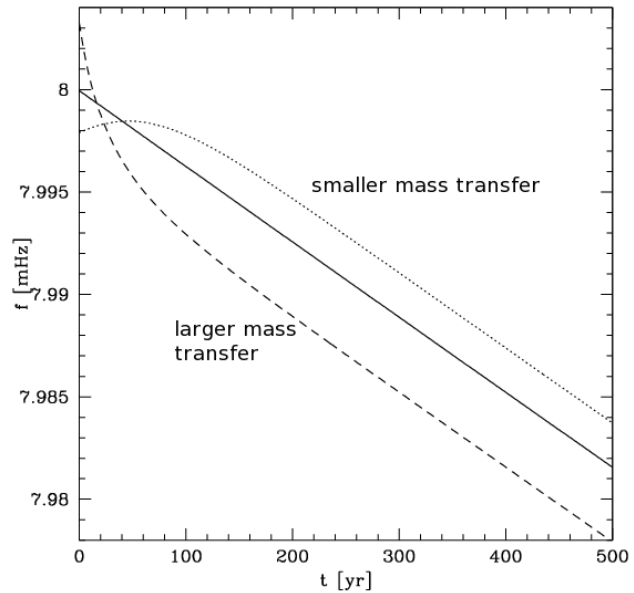}}
\caption[The evolution of the gravitational wave frequency in a
  perturbed AM CVn system over time.]{The evolution of the
  gravitational wave frequency {of the fiducial AM CVn system} over time, with
the x axis counting from the beginning of the perturbation. The solid line
denotes the equilibrium evolution, the dotted line an evolution with a mass
transfer rate five times smaller than in equilibrium at onset of
perturbation, the dashed line an evolution with a mass transfer rate five
times bigger than in equilibrium at onset of perturbation. 
}
\label{short1_1}
\end{figure}

In Fig.~\ref{short1_1} we show the frequency evolution of the unperturbed system (solid line) and the perturbed system (dashed and dotted lines).
We observed a change in drift of the frequency of gravitational wave
emissions after a perturbation struck the system. Perturbations that
lead to increased mass transfer rates forced the system to separate at
a higher pace, leaving the gravitational wave frequency to decrease at
a faster pace as compared to a secular evolving AM CVn at the same
stage of evolution ({compare} Fig.~\ref{short1_1} dashed
line {with the} solid
line). Perturbations that decreased the mass transfer rate, in contrast,
lead to a short in-spiral of the system over 59 years (see
Fig.~\ref{short1_1} dotted line).  {The system experiences a new ``turning point'' and
  then evolves to lower frequencies again.}

\begin{figure}
\resizebox{\hsize}{!}{\includegraphics{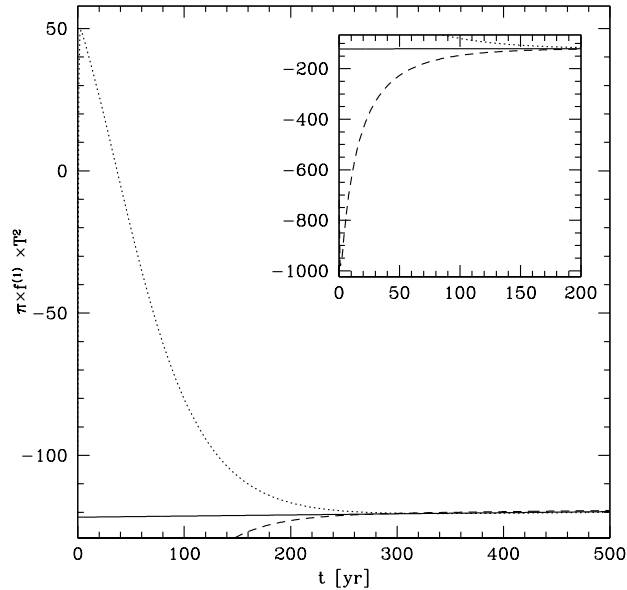}}
\caption[Number of wave-cycles as contributed by the first derivative
of the gravitational wave frequency (spin-down parameter) in perturbed
AM CVn systems over time as seen by LISA]{Number of wave-cycles as
contributed by the first derivative of the gravitational wave
frequency (spin-down parameter) over time as seen by LISA during a five
year observation. {Time as shown on the x axis refers to the timeline since perturbation struck the system, 
and a hypothetical LISA observation is started at every instance of this timeline to yield shown wavecycle evolution.} 
The solid line denotes the equilibrium evolution, the
dotted line an evolution with a mass transfer rate
{five} times smaller than in equilibrium at onset of
perturbation, the dashed line an evolution with a mass transfer rate
{five} times bigger than in equilibrium at onset of
perturbation. The inlay shows a zoom in in order to cover the full
spread of the evolution with increased mass transfer rate at
onset. For details see text.}
\label{short2}
\end{figure}

\begin{figure}
\resizebox{\hsize}{!}{\includegraphics{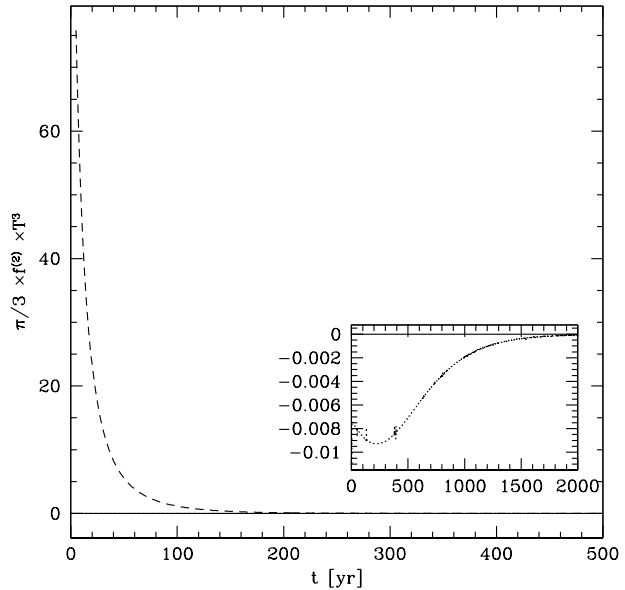}}
\caption[Number of wave-cycles as contributed by the second derivative
of the gravitational wave frequency (spin-down parameter) in perturbed
AM CVn systems over time as seen by LISA]{Number of wave-cycles as
contributed by the second derivative of the gravitational wave
frequency (spin-down parameter) over time as seen by LISA during a five
year observation. Conventions as in Fig.~\ref{short2}}
\label{short3}
\end{figure}

We determined the total number of wave-cycles as detected during a one
and five year mission of LISA and the contributions from the spin-down
parameters to this total number by ``starting'' an observation at each
individual year of the shown frequency evolution of the AM CVn in
Fig.~\ref{short1_1} (see Fig.~\ref{short2} and Fig.~\ref{short3}). We
found, that the shapes of {the} curves in
Fig.~\ref{short2} and Fig.~\ref{short3} do not
{change} between a one year observation and a five
year observation, they simply scale according to {$T^2$ and $T^3$ for the first and second derivative
respectively}. We therefore show explicit examples for a 5 year mission only.

The number of wave cycles contributed by $f_0^{(1)}$ clearly mirrors
the altered frequency evolution. Perturbations that force increased
mass transfer rates (dashed lines) yield faster separating systems, {i.e.} large
negative contributions by $f_0^{(1)}$. We find  the number of
wave cycles from the first spin-down parameter to increase from -123
wave cycles to -990 wave cycles when integrating the number of
wave cycles over a five year observation starting an instant after the
sudden offset of $\Delta$. Perturbations inducing decreased mass
transfer rates (dotted lines) yield a short in-spiral of the system, we thus see first
moderate positive then moderate negative contributions by $f_0^{(1)}$
as separated by the new turning point in frequency. Here the
significant alteration is more the reverse of the drift of the system,
with +50 wave cycles immediately after the impulse instead of the
quoted -123 wave cycles.

Interesting results can be found for $f_0^{(2)}$. This spin-down
parameter {normally does not contribute} to the total number of
wave-cycles {for a system at 8mHz} even in a five year
observation. In case of a faster separating system however we suddenly
see a significant increase in the contribution to the total number of
wave-cycles, in fact by a factor of $10^6$. Suddenly the second
derivative becomes visible within a five year mission with +76
additional contributed wave-cycles, compared to a null contribution
beforehand. In case of the  system {with
decreased mass transfer} we see an increase in the number of
contributed wave cycles, but still the contribution is too small to be
{detected}. 

\section{Discussion and conclusions}\label{sec:disc}

{We studied the reaction of the frequency evolution of a fiducial
  AM CVn system with a GW frequency of 8mHz, to sudden, but modest,
  perturbations of its orbital parameters. These perturbations are
  inspired by the observed and theoretically expected short term
  variations due to e.g. nova explosions. The system remains bound,
  changes its mass transfer rate, but on a time scale of hundreds of
  years, evolves back to its equilibrium mass transfer rate, as
  expected. However, if any of the thousands of individual binaries
  that LISA is expected to follow during its life time, is in such a
  perturbed state, the measured frequency evolution can be
  dramatically different than for the equilibrium case. The sign of
  the frequency derivative may change and the number of cycles
  contributed by the first and second derivative may increase by
  several tens to several hundreds in a 1 or 5 year mission life
  time. For template data analysis methods, the detection of
  these systems should still be secure, if sufficient spin-down
  parameters are used (up to second derivative, especially for a 5 year
  mission). However, care should be taken with deriving system
  parameters purely on the GW data, as the perturbations can give rise
  to wildly different derivatives, which, if interpreted as
  equilibrium evolution, would lead to extreme mass estimates, or for
  the systems with reversed evolution to the conclusion that no mass
  transfer is present. In addition in \citet{Stroeer2005a} we
  concluded that $f_0^{(2)}$ is needed for distance estimations, but
  this works only when the system evolves in equilibrium.}

{It is difficult to estimate the number of LISA binaries that are affected by 
such short term variations. Many observed interacting binaries seem to have 
short term period evolution different from the secular evolution (e.g. 
\citet{2003MNRAS.345.889B}), suggesting it may be a quite common phenomenon. For 
the helium novae in AM CVn systems we can make a simple estimate. At the 
shortest periods, i.e. high mass transfer rates and for high accretor masses 
the amount of accumulated mass needed to ignite the nova is about 0.001 Msun 
\citep{2007ApJ...662L..95B}. For a mass transfer rate of $10^{-5}$ Msun/yr, this 
leads to a nova every 100 years, influencing the evolution essentially all 
the time so we expect several LISA sources to be affected. However, most of 
the LISA sources will have evolved to lower mass transfer rates, where the 
nova frequency is much lower and the chances of catching a system in the 
recovery phase are low.}

{We also tested repercussions of short term perturbations at an 
evolutionary stage of the fiducial AM CVn at which the orbital 
separation yields GW frequencies of order 3 mHz. This stage is set 3.5 
million years after we found our AM CVn at 8mHz. Overall, tidal coupling 
is much less at such a separation so that the response to a shortterm 
perturbation is less dynamic. We find the timescale of the perturbation, 
$\tau_\mathrm{c}$, to be 691 yrs compared to 47.9 yrs. Therefore the maximum 
amount of increase in wavecycles, here found in the increased mass 
transferring scenario, is only a mere 2.5 wavecycles, and the second 
derivative is never visible in case of increased and decreased mass 
transfer perturbations at proposed scale. However, decreased mass 
transfer is still able to trigger a new turning point in the frequency 
evolution of the binary, the system inspirals again over a period of 
roughly 2000 years. We therefore conclude that although the magnitude of 
wavecycle increase is minor at 3 mHz the actual ability of the system to 
shortly inspiral is a reason of concern once again for parameter 
inference.}

{We thus re-enforce our earlier conclusion \citep{Stroeer2005a}
  that complementary electro-magnetic data from e.g. GAIA \citep{2009CQGra..26i4030N} is an
  important ingredient for the disentangling of different processes
  that may contribute to the frequency evolution of compact binaries,
  such as mass transfer, tidal interaction and short term variability
  and the determination of accurate system parameters.}

\bibliographystyle{mn2e.bst}
\bibliography{intro,rjmcmc,references,lisa,a,ref}

\end{document}